\documentclass{ws-ijmpa}

\begin{document}

\markboth{G.~L.~Klimchitskaya {\it et al.}}
{Casimir Effect as a Test for Hypothetical Interactions}

%
\catchline{}{}{}{}{}
%

\title{CASIMIR EFFECT AS A TEST FOR THERMAL CORRECTIONS \\
AND HYPOTHETICAL
LONG-RANGE INTERACTIONS\footnote{This paper is the combined
presentation of two talks given by G.\ L.\ Klimchitskaya and
by V.\ M.\ Mostepanenko.}
}

\author{G.~L.~KLIMCHITSKAYA}

\address{North-West Technical University,\\ Millionnaya St.\ 5, St.Petersburg,
191065, Russia\\
Galina.Klimchitskaya@itp.uni-leipzig.de
}

\author{R.~S.~DECCA}

\address{Department of Physics, Indiana
University-Purdue
University Indianapolis,\\
Indianapolis, Indiana 46202, USA
}

\author{E.~FISCHBACH}

\address{Department of Physics, Purdue University,\\
West Lafayette, Indiana 47907, USA
}

\author{D.~E.~KRAUSE}

\address{Physics Department, Wabash College, Crawfordsville,\\
Indiana 47933, USA
}

\author{D.~L\'{O}PEZ}

\address{Bell Laboratories, Lucent Technologies,\\ Murray Hill, New
Jersey 07974, USA
}

\author{V.~M.~MOSTEPANENKO}

\address{Noncommercial
Partnership ``Scientific Instruments'',\\
Tverskaya St.\ 11, Moscow, 103905,  Russia\\
\smallskip
and \\
\smallskip
A.~Friedmann Laboratory, St.Petersburg, Russia
}

\maketitle

\pub{Received 16 November 2004}{}

\begin{abstract}
We have performed a precise experimental determination of the Casimir pressure
between two gold-coated parallel plates by means of a micromachined
oscillator. In contrast to all previous experiments on the Casimir effect,
where a small relative error (varying from 1\% to 15\%) was achieved
only at the shortest separation, our smallest experimental error
($\sim 0.5$\%) is achieved over a wide separation range
from 170\,nm to 300\,nm at 95\%
confidence. We have formulated a rigorous metrological procedure for the
comparison of experiment and theory without resorting to the previously used
root-mean-square deviation, which has been
criticized in the literature. This enables us to discriminate among
different competing theories of the thermal Casimir force,
and to resolve a thermodynamic puzzle arising from the
application of Lifshitz theory
to real metals. Our results lead to a more rigorous approach
for obtaining constraints on hypothetical long-range interactions
predicted by extra-dimensional physics and other extensions of the
Standard Model. In particular, the constraints on non-Newtonian
gravity are strengthened by up to a factor of 20 in a wide interaction
range at 95\% confidence.

\end{abstract}

\section{Introduction}

The Casimir force\cite{1} is a macroscopic quantum phenomenon arising
from the modification of the zero-point oscillations of
the electromagnetic field
by material boundaries. It is in fact the limiting case of the well
known van der Waals force when the separation distances between the
test bodies are large enough for the manifestation of retardation
effects. The modern experimental study of the Casimir force began
in 1997 with Ref.~\refcite{2} followed by 
Refs.~\refcite{3}--\refcite{11a}.
This work was stimulated both by the demands of nanotechnology,
where the Casimir force
may be large enough to drive nanoscale
devices,\cite{8} and also by the theory
of fundamental interactions, where the Casimir effect plays a large role
in testing the predictions of extra-dimensional physics.\cite{11a}
Simultaneously with these new experiments, extensive theoretical work was
carried out to refine calculations of the Casimir force by taking into account
numerous correction factors such as finite conductivity of the boundary
metal, nonzero temperature, surface roughness, and patch potentials
among others (see Ref.~\refcite{12} for a review).

Beginning in 2000, the behavior of the thermal correction to the
Casimir force between real metals has been hotly debated. It was shown
that  Lifshitz theory,\cite{13} which provides the theoretical
foundation for the calculations of both the van der Waals and Casimir
forces, leads to different results depending on the model of metal
conductivity used. For real metals at low frequencies $\omega$,
the dielectric permittivity $\varepsilon$ varies as $\omega^{-1}$. After
substituting $\varepsilon\sim\omega^{-1}$
 into the Lifshitz formula, the result is a thermal
correction which is several hundred times greater than for
ideal metals at separations of a few tenths of a micrometer.\cite{14,15}
The attempt\cite{16} to modify the zero-frequency term of the Lifshitz
formula for real metals, assuming that it behaves as in the case of
ideal metals, also leads
to a large thermal correction to the Casimir force at short separations.

It is important to note
that in the
approaches of both Refs.~\refcite{14,15} and also of 
Ref.~\refcite{16} a thermodymanic puzzle
arises, i.e., the Nernst heat theorem is violated for a perfect
lattice.\cite{17,17a} (See also Ref.~\refcite{17b} where it is shown that for
the preservation of the Nernst heat theorem in the approach of
Refs.~\refcite{14},\ \refcite{15} it is necessary to have metals with
defects or impurities; it is common knowledge, however, that thermodynamics
must be valid for both perfect and imperfect lattices.)
This puzzle casts doubt on the many applications of the Lifshitz theory
of dispersion forces, and thus represents a potentially serious challenge to
both experimental and theoretical physics. By contrast, the use of
$\varepsilon\sim\omega^{-2}$, as holds in a free electron plasma
model neglecting relaxation, leads\cite{18,19} to a small thermal 
correction to the Casimir force at short separations.
This is in qualitative agreement with
the case of an ideal metal and is consistent with the Nernst heat
theorem. It should be borne in mind, however, that the plasma model
is not universal, and is applicable only in the case when the characteristic
frequency is in the domain of infrared optics.

A universal theoretical approach consistent with thermodynamics was proposed
in Ref.~\refcite{20}. 
It uses the Lifshitz formula with reflection coefficients
expressed in terms of the surface impedance instead of the dielectric
permittivity. In the framework of this approach one need not
consider the zero-point and thermal photons inside a metal. The impedance
approach was found to be consistent with the experimental
results of Ref.~\refcite{11,11a},
whereas the alternative approaches of Refs.~\refcite{14},\ \refcite{15} 
and Ref.~\refcite{16}
were excluded by this experiment. It should be noted, however, that in
Refs.~\refcite{11},\ \refcite{11a} the contribution of surface 
roughness was rather large, 
and theory was compared with experiment by the use of the root-mean-square
deviation (a method previously used in experiments on the Casimir force
and criticized in literature\cite{7}).

In the present paper we carry out a
precise experimental determination of the Casimir pressure
between two gold-coated parallel plates by means of a micromachined
oscillator. In contrast to all previous experiments on the Casimir force,
the smallest experimental relative error (in our case from 0.55\% to 0.6\%
at 95\% confidence)
was achieved not only at the shortest separations but rather
in a wide separation range from 170\,nm to 300\,nm.
The theoretical values of the Casimir pressure were calculated in the
framework of each of the aforementioned approaches,
taking careful account of all
relevant corrections. The error in the theoretical results was found 
independently of the experimental errors. The distinguishing feature of
our present comparison of experiment with theory is that we do not use the
root-mean-square deviation. Instead, a rigorous
metrological procedure is applied, which permits us to conclusively
exclude the alternative approaches of Refs.~\refcite{14}--\refcite{16}
to the thermal Casimir force, and to thus
resolve the puzzle arising from the violation of the Nernst heat theorem
in these approaches.
Our results are used to strengthen constraints on
non-Newtonian gravity in the micrometer range by a factor of up to 20, 
 and to significantly increase their reliability.

\section{Setup, Measurement Procedure and Experimental Errors}

\begin{figure}[t]
\vspace*{-7cm}
\centerline{\psfig{file=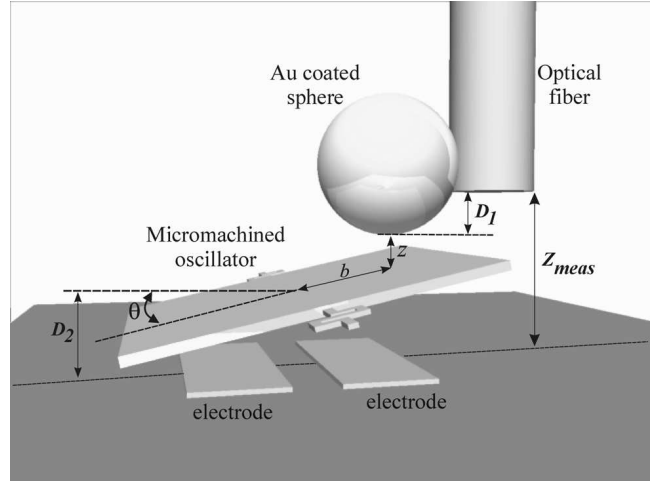,width=20cm}}
\vspace*{-11cm}
\caption{Schematic of the experimental
setup.}
\end{figure}

The details of the experimental setup containing the micromachined
oscillator (see Fig.~1) have already been presented in 
Refs.~\refcite{11},\ \refcite{11a}.
The vertical separation $z$ between the sapphire sphere of radius
$R=(148.7\pm 0.2)\,\mu$m and a $500\times 500\,\mu\mbox{m}^2$ heavily
doped polysilicon plate (both coated with gold) was varied harmonically
in time at the natural resonant frequency of the micromachined
oscillator $\omega_o=2\pi\times702.92\,$Hz. Under the influence of
the Casimir force $F(z)$ acting between a plate and a sphere, the
resonant frequency shifts, and from the measurement of this shift
one can calculate $\partial F(z)/\partial z$ (see
Refs.~\refcite{8},\ \refcite{11}).
Using the proximity force theorem,\cite{12} we can then find the equivalent
Casimir pressure
\begin{equation}
P(z)=-\frac{1}{2\pi R}\frac{\partial F(z)}{\partial z}
\label{eq1}
\end{equation}
\noindent
between the two parallel plates, i.e., the Casimir force per unit area.

In the present experiment the setup was significantly improved
in several ways, which
permitted us to obtain results which considerably
exceed those of Refs.~\refcite{11},\ \refcite{11a} in
consistency and conclusiveness.
First, the surface roughness was decreased by an order of magnitude on
the sphere, and by a factor of five on the plate. This resulted in maximal
heights of the roughness peaks equal to 11.06\,nm and 20.63\,nm,
respectively. For the roughness characterization, the samples were
studied with an AFM both before and after the Casimir force measurements.
Second, the error in the measurements of the absolute separation $z$
between the sphere and the plate was decreased from $\Delta z=1\,$nm
to $\Delta z=0.6\,$nm at 95\% confidence. The absolute separations
were determined from $z=z_{meas}-D-b\theta$ (see Fig.~1), where
$D=D_1+D_2$ and the lever arm $b=(210\pm 3)\,\mu$m. The quantity
$z_{meas}$ is measured interferometrically, and $\theta$ is determined
from the difference in capacitance between the plate and the right and left
underlying electrodes. The value of $D=(9349.7\pm 0.5)\,$nm was
found from 120 plots of the electrostatic force as functions of
separation
at $z>3\,\mu$m (where the Casimir force is negligible) for the
given potential differences between a sphere and a plate.
Third, shorter separation distances between the sphere and the
plate were achieved (160\,nm instead of 260\,nm) owing to an
improvement in detection sensitivity, and to a decrease of the coupling
between the micromachined oscillator and the environment. In doing so
we verified that the response of the micromachined oscillator
was still linear.
Fourth, a gold coating was used on both test bodies (instead of the
dissimilar metals as in Refs.~\refcite{11},\ \refcite{11a}) 
which makes the theoretical
interpretation of the final results more transparent.
As a result, the Casimir pressure $P^{\rm expt}(z)$ was measured
within a separation range from 160\,nm  to 750\,nm. This measurement was
repeated fifteen times with 288--293 points in each run. Each
individual point was obtained with an integration time of 10\,s.

The  experimental data were analyzed for the presence of
outlying results by the use of statistical criteria,\cite{21} and
one of the fifteen sets of measurements was found to be outlying.
All data from the remaining fourteen sets were plotted as a function of
separation for separation distances between 160\,nm
and 750\,nm, where the total separation interval was divided
into partial subintervals of length $2\Delta z=1.2\,$nm each. 
The measurement data were carefully analyzed and
found to be uniform in mean values, but not uniform in variance.
Because of this, the random error as a function of separation was
found using a special procedure developed from the theory of repeated
measurements.\cite{22,23}

\begin{figure}[t]
\vspace*{-5.5cm}
\centerline{\psfig{file=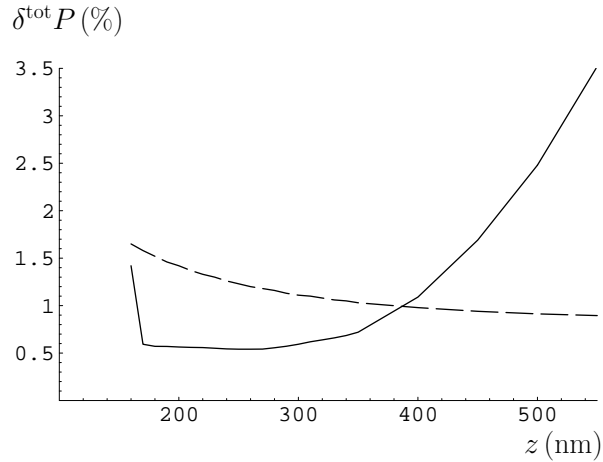,width=14cm}}
\vspace*{-6cm}
\caption{The total relative errors of our experimental
(solid line) and theoretical (dashed line)
Casimir pressures versus separation.
}
\end{figure}

The systematic errors in the Casimir pressure determination were found
to arise from the error in sphere radius $\Delta R=0.2\,\mu$m, and
from the error in the angular resonant frequency
$\Delta\omega_o=2\pi\times 6\,$mHz. 
To obtain the total experimental error of our measurements,
it is necessary to combine the random and systematic errors which are
described by the normal or Student and uniform distributions,
respectively. Finally, the total relative
experimental error as a function of separation was determined at 95\%
confidence (see solid line in Fig.~2). As is seen from Fig.~2,
the smallest experimental error from 0.55\% to 0.6\% is achieved, not
only at the shortest separation (as in all previous Casimir force
measurements where the error was also larger), but in a wide range from
170\,nm to 300\,nm. This opens new opportunities to use our
results for imposing stronger constraints on thermal effects and on 
hypothetical long-range interactions.

\section{Theory, Theoretical Errors and Comparison with Experiment}

In addition to performing the measurements, we calculated
the equivalent Casimir
pressure acting between two gold-coated plates in thermal equilibrium
at a temperature $T=300\,$K using the
Lifshitz formula\cite{13}
\begin{eqnarray}
&&
P(z)=-\frac{k_BT}{\pi}\sum\limits_{l=0}^{\infty}{\vphantom{\sum}}^{\prime}
\int_0^{\infty}k_{\bot}dk_{\bot}q_l
\label{eq2} \\
&&
\phantom{aaa}
\times\left\{\left[r_{\|}^{-2}(\xi_l,k_{\bot})e^{2q_lz}-1\right]^{-1}+
\left[r_{\bot}^{-2}(\xi_l,k_{\bot})e^{2q_lz}-1\right]^{-1}\right\}.
\nonumber
\end{eqnarray}
\noindent
Here $q_l^2=k_{\bot}^2+\xi_l^2/c^2$ and
the summation is performed with respect to the Matsubara frequencies
$\xi_l=2\pi k_BTl/\hbar$, where $k_B$ is the Boltzmann constant,
$l=0,\,1,\,2,\,\ldots\,$, and the prime adds a factor 1/2 for
the $l=0$ term. The reflection coefficients
$r_{\|,\bot}$ also depend on the magnitude of a wave vector
component in the plane of plates
$k_{\bot}=|\mbox{{\boldmath{$k$}}}_{\bot}|$.

As was shown in Ref.~\refcite{20}, in order for the Lifshitz formula
to avoid contradictions with thermodynamics for real metals,
the reflection coefficients should be expressed in terms of the
surface impedance $Z(\omega)$:
\begin{equation}
r_{\|}^{-2}(\xi_l,k_{\bot})=\left[
\frac{Z(i\xi_l)\xi_l+cq_l}{Z(i\xi_l)\xi_l-cq_l}\right]^2,
\quad
r_{\bot}^{-2}(\xi_l,k_{\bot})=\left[
\frac{Z(i\xi_l)cq_l+\xi_l}{Z(i\xi_l)cq_l-\xi_l}\right]^2.
\label{eq3}
\end{equation}
\noindent
The values of the impedance at all contributing imaginary Matsubara
frequencies with $l\geq 1$ are given by
$Z(i\xi_l)=1/\sqrt{\varepsilon(i\xi_l)}$ (which is the so called
Leontovich impedance\cite{23a}), where $\varepsilon(i\xi_l)$
is calculated by means of a dispersion relation from the tabulated\cite{24}
complex refraction index of Au (see, e.g., Ref.~\refcite{25} for
the details of the calculation procedure). When the Matsubara
frequency is zero, the value of the impedance should be obtained\cite{20}
by extrapolation from the region of characteristic frequencies,
which is infrared optics in our case, resulting in
$Z(i\xi)\approx \xi/\omega_p$ when $\xi\to 0$, where $\omega_p$ is
the plasma frequency. From Eq.~(\ref{eq3})
one then arrives at
\begin{equation}
r_{\|}^{-2}(0,k_{\bot})=1, \qquad
r_{\bot}^{-2}(0,k_{\bot})=
\left(\frac{ck_{\bot}+\omega_p}{ck_{\bot}-\omega_p}\right)^2.
\label{eq4}
\end{equation}

The use of the Leontovich impedance in Eq.~(\ref{eq3}) which does not depend
on the polarization state and transverse momentum, is of prime importance.
Note that in Refs.~\refcite{25a},\ \refcite{25b} the exact impedances
depending on a transverse momentum were used. This has led to the
same conclusions as were obtained previously from the Lifshitz formula
combined with the dielectric permittivity $\varepsilon\sim\omega^{-1}$.
As was already mentioned above, these conclusions are in violation
of the Nernst heat theorem for a perfect 
lattice.\cite{17}${}^{-}$\cite{17b}
Although a recent review\cite{25b} claims agreement 
with the Nernst heat theorem
in Refs.~\refcite{14},\ \refcite{15},  no specific objections
against the rigorous analytical proof of the opposite statement in
Ref.~\refcite{17a} are presented.
The fallacy in the calculations of Refs.~\refcite{25a},\ \refcite{25b}
concerning the type of the impedance
is that they disregard the requirement that the reflection properties
for virtual photons on a classical boundary should be the same as
for real photons. Ref.~\refcite{17a} demonstrates
in detail that by enforcing this requirement the exact and Leontovich
impedances coincide at zero frequency and lead to the conclusions
of Ref.~\refcite{20} which are in perfect agreement with the Nernst
heat theorem. 

In fact, the exact impedances, depending on both polarization state
and transverse momentum are\cite{17a,25a,25b}
\begin{eqnarray}
&&
Z_{\|}(\omega,k_{\bot})=\frac{1}{\sqrt{\varepsilon(\omega)}}
\left[{1-\frac{c^2k_{\bot}^2}{\varepsilon(\omega)\omega^2}}
\right]^{1/2},
\nonumber \\
&& \label{eq4a} \\
&&Z_{\bot}(\omega,k_{\bot})=\frac{1}{\sqrt{\varepsilon(\omega)}}
\left[{1-\frac{c^2k_{\bot}^2}{\varepsilon(\omega)\omega^2}}
\right]^{-1/2}.
\nonumber
\end{eqnarray}
\noindent
The angle of incidence $\theta_0$ of a plane electromagnetic wave
with a wave vector {\boldmath$k$} from vacuum on the boundary plane of
a metal is evidently given by 
$\sin\theta_0=k_{\bot}/|{\mbox{\boldmath$k$}}|$.
The important requirement is that the reflection properties of virtual
photons on a classical boundary are the same as of real ones.
In particular, the angle of reflection must be equal to the angle
of incidence, i.e., the first Snell's law must be satisfied. This law
follows from the fact that for reflection at a classical 
boundary, the mass-shell equation 
$|{\mbox{\boldmath$k$}}|=\omega/c$ is valid leading to 
$\sin\theta_0=ck_{\bot}/\omega$. Substituting this into Eq.~(\ref{eq4a})
and expanding in powers of a small parameter $\sin^2\theta_0/\varepsilon$,
we arrive at
\begin{eqnarray}
&&Z_{\|}(\omega,k_{\bot})=\frac{1}{\sqrt{\varepsilon(\omega)}}
\left[{1-\frac{\sin^2\theta_0}{\varepsilon(\omega)}}\right]^{1/2}
=Z(\omega)\left[1-\frac{\sin^2\theta_0}{2\varepsilon(\omega)}
+\ldots\right],
\nonumber\\
&&\label{eq4b}\\
&&Z_{\bot}(\omega,k_{\bot})=\frac{1}{\sqrt{\varepsilon(\omega)}}
\left[{1-\frac{\sin^2\theta_0}{\varepsilon(\omega)}}\right]^{-1/2}
=Z(\omega)\left[1+\frac{\sin^2\theta_0}{2\varepsilon(\omega)}
-\ldots\right],
\nonumber
\end{eqnarray}
\noindent
where $Z(\omega)$ is the Leontovich impedance.

In metals for all frequencies which are at least several times 
smaller than the plasma frequency $|\varepsilon|\gg 1$ holds. 
For this reason, the term $\sin^2\theta_0/\varepsilon$ in 
Eq.~(\ref{eq4b}) can be neglected in comparison with unity. 
An important point is that $|\sin^2\theta_0/\varepsilon|\to 0$
when $\omega\to 0$, i.e., at zero frequency the Leontovich impedance
precisely coincides with the exact impedances. This is
in contradiction with Refs.~\refcite{25a},\ \refcite{25b}, where
the limit $\omega\to 0$ was considered at fixed nonzero $k_{\bot}$.
The latter violates the mass-shell equation and, consequently, 
necessitates
postulating some unusual reflection properties for virtual photons on 
a classical boundary. As a result, the use of the Leontovich impedance,
which is in fact exact at zero frequency, is in agreement with 
thermodynamics, whereas the approach of 
Refs.~\refcite{25a},\ \refcite{25b} leads to the previous
conclusions\cite{15} 
which are in contradiction with fundamental physical principles. 

The claims of Ref.~\refcite{25b} against the
extrapolation of Eq.~(\ref{eq4}) to the zero Matsubara frequency
also collapse.\cite{25c} As is shown in Ref.~\refcite{25c}, the
alternative extrapolation, suggested in Ref.~\refcite{25b}, leads
to the unsupportable violation of the Nernst heat theorem.

Using Eqs.~(\ref{eq2})--(\ref{eq4}), the Casimir pressures  $P(z)$
were computed within the measurement separation range. 
It can be seen\cite{6a}
that small sample-to-sample variations of the tabulated data for the
refractive index due to, for example, differences in the grain size
distribution and the presence of impurities, lead to
a decrease in the magnitude of the Casimir pressure which
is much less than 0.5\%
even at the shortest separation $z=160\,$nm considered here.
Note also that at separations $160\,\mbox{nm}\leq z\leq 750\,$nm
the effects of spatial dispersion do not lead to any noticeable
contribution to the Casimir pressure.\cite{26}

The other correction factors which may influence the values of the
theoretical Casimir pressure are the surface roughness and patch
potentials. The AFM study of both interacting surfaces shows that
the characteristic lateral size of the surface distortions is approximately
500--600\,nm, i.e., a factor of three larger than the shortest
separations in our experiment. In this case the roughness correction
can be found\cite{27} by  geometrical averaging,\cite{3a,12}
which leads to the
conclusion that the largest roughness correction achieved here at the
shortest separation ($z=160\,$nm) is only 0.65\% of the Casimir pressure
(and 0.42\% at $z=200\,$nm). It is easily seen that the contribution
of diffraction-type and correlation effects to the roughness
correction, which cannot be found by the additive method, is of order of
0.01\%. The contributions of patch potentials, which are of the same
order, and thus negligible, were found in analogy with Ref.~\refcite{6a}.
The values of the Casimir pressure $P^{\rm theor}(z)$
including roughness corrections were computed at all 4066 experimental
points.

We are now in a position to determine the error in the theoretical
results. The main sources of errors are the proximity force theorem
[see Eq.~(\ref{eq1})], which leads\cite{28}${}^{-}$\cite{30} 
to a relative error in the Casimir pressure less
than $z/R$, and an uncertainty in the tabulated data
for the complex refractive index leading to an error less than 0.5\%.
Both of these errors in the pressure are approximately described by
a uniform distribution.
One should also take into account that, for purposes of comparison with
experiment, the theoretical pressure is computed at the experimental points
defined with an error $\Delta z$.
Since the Casimir pressure depends on the inverse fourth power of the
separation, this leads\cite{31} to a relative error $4\Delta z/z$.
Combining the above three errors at 95\% confidence, we obtain the
theoretical relative error of the Casimir pressure calculations shown by
the dashed line in Fig.~2. As is seen from Fig.~2, at separations
$z<390\,$nm the experimental error is less than the theoretical
error.

\begin{figure}[t]
\vspace*{-5.2cm}
\centerline{\psfig{file=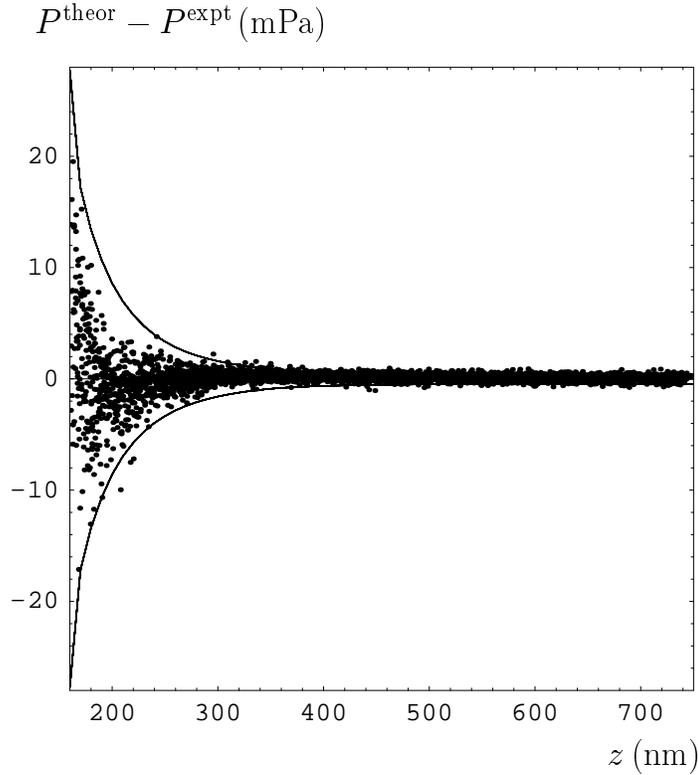,width=16cm}}
\vspace*{-6.5cm}
\caption{\label{expthKM}
Differences of our theoretical and experimental Casimir pressures
(dots) and 95\% confidence interval (solid lines) versus
separation for all fourteen sets of measurements.}
\end{figure}

With independently determined experimental and theoretical errors at our
disposal, it becomes possible to find the total absolute error of the new
random variable $\left[P^{\rm theor}(z)-P^{\rm expt}(z)\right]$
at 95\% confidence.
In this way we avoid the use of the root-mean-square deviation between
theory and experiment which was applied in previous Casimir force measurements.
In Ref.~\refcite{7} this procedure was demonstrated to be inadequate
when the force
increases rapidly with the decrease of separation, although no
alternative approach was proposed. In Fig.~\ref{expthKM} 
the confidence interval
$\left[-\Delta^{\!{\rm tot}}(P^{\rm theor}-P^{\rm expt}),
\Delta^{\!{\rm tot}}(P^{\rm theor}-P^{\rm expt})\right]$ is shown by the
solid lines as a function of separation. In the same figure the differences
between our theoretical and experimental Casimir
pressures are plotted by points
for all fourteen sets of measurements. Remarkably, only 207 points
(i.e., 5.09\% of the total number) fall outside of the confidence
interval, which demonstrates excellent agreement between our theory and
experiment. The relative measure of agreement between theory and
experiment is
$\Delta^{\!{\rm tot}}(P^{\rm theor}-P^{\rm expt})/|P^{\rm theor}|$.
This is equal to 1.9\% at $z=170\,$nm, 1.4\% within the interval
$270\,\mbox{nm}\leq z\leq 370\,$nm, 1.8\% at $z=420\,$nm, and then
increases up to 13\% at $z=750\,$nm.
Thus our experiment is the first measurement of the Casimir force
where agreement between theory and experiment 
at the level of 1.5\% has been achieved at high confidence
within a wide separation region.

\section{Alternative Theories and Resolution of a Thermodynamic
Puzzle}

\begin{figure}[t]
\vspace*{-5.cm}
\centerline{\psfig{file=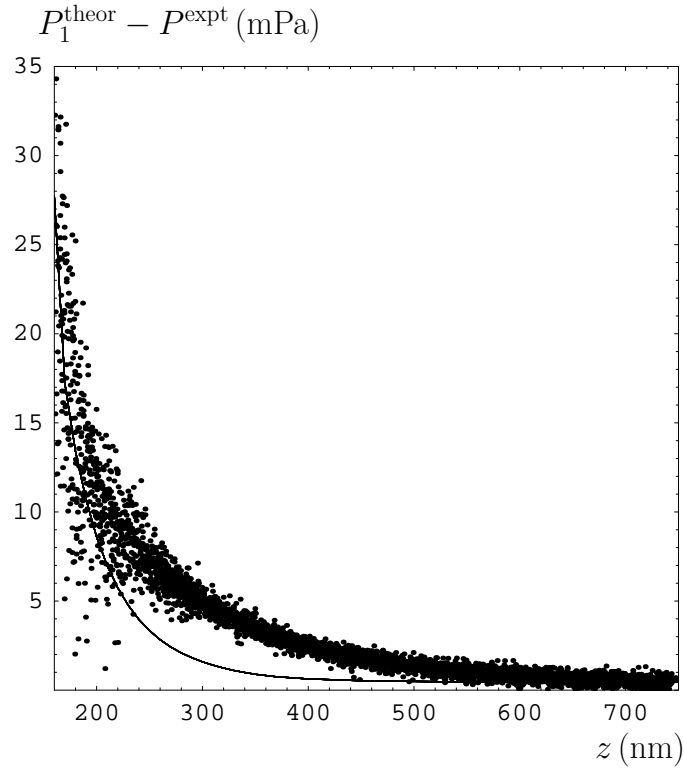,width=16cm}}
\vspace*{-5.cm}
\caption{\label{expthBS} Differences of theoretical${}^{17,18}$ 
and experimental Casimir pressures
(dots) and a positive half of a 95\% confidence interval (solid line) versus
separation for all fourteen sets of measurements.}
\end{figure}
\begin{figure}[t]
\vspace*{-5.cm}
\centerline{\psfig{file=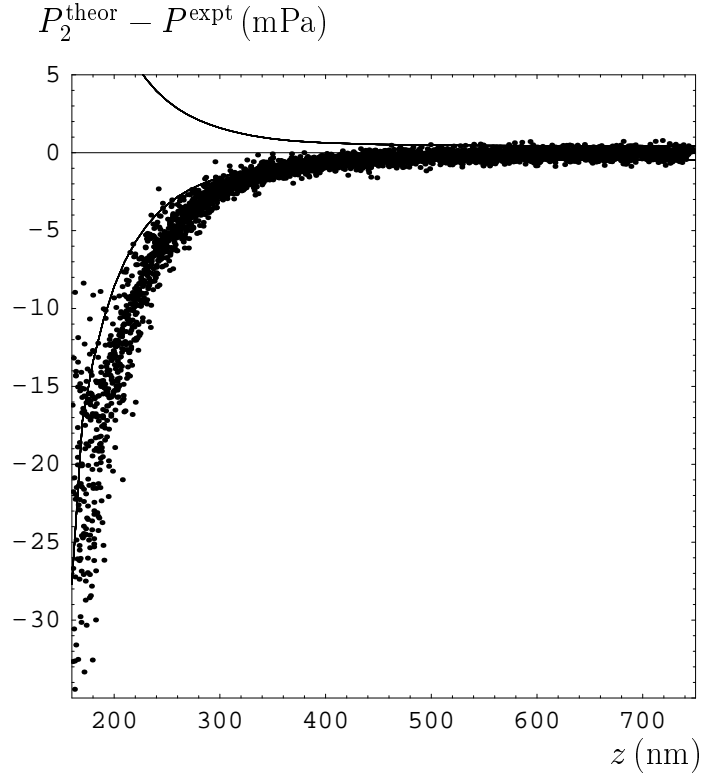,width=16cm}}
\vspace*{-6.5cm}
\caption{\label{expthSL} Differences of theoretical${}^{19}$
and experimental Casimir pressures
(dots) and 95\% confidence interval (solid lines) versus
separation for all fourteen sets of measurements.}
\end{figure}
This experiment, and the extent of agreement between our measurements
and theory, can be used to test the alternative theoretical approaches
to the thermal Casimir force
proposed in Refs.~\refcite{14}--\refcite{16}, 
and to thus finally resolve the
contradiction between these approaches and thermodynamics.
Refs.~\refcite{14}--\refcite{16}
use the Lifshitz formula in Eq.~(\ref{eq2}) with the reflection coefficients
expressed in terms of the dielectric permittivity
\begin{equation}
r_{\|,L}^{-2}(\xi_l,k_{\bot})=\left[
\frac{k_l+\varepsilon(i\xi_l)q_l}{k_l-\varepsilon(i\xi_l)q_l}\right]^2,
\quad
r_{\bot,L}^{-2}(\xi_l,k_{\bot})=\left[
\frac{k_l+q_l}{k_l-q_l}\right]^2,
\label{eq5}
\end{equation}
\noindent
where $k_l^2=k_{\bot}^2+\varepsilon(i\xi_l)\xi_l^2/c^2$.
The values of $\varepsilon(i\xi)$ in 
Refs.~\refcite{14}--\refcite{16} were found
in the same way as described above. In fact the use of the Lifshitz
reflection coefficients (\ref{eq5}) instead of the impedance
coefficients (\ref{eq3})
leads to only minor differences at all $l\geq 1$. The major difference
in the approaches of Refs.~\refcite{14}--\refcite{16} is the value of the
zero-frequency term of the Lifshitz formula. 
In Refs.~\refcite{14},\ \refcite{15} 
(see also Refs.~\refcite{25a},\ \refcite{25b})
the Drude dielectric function, depending on frequency as
$\varepsilon\sim\omega^{-1}$ when $\omega\to 0$, was substituted into
Eq.~(\ref{eq5}) leading to $r_{\|,L}^2(0,k_{\bot})=1$,
$r_{\bot,L}^2(0,k_{\bot})=0$ instead of Eq.~(\ref{eq4}).
In Ref.~\refcite{16} it was postulated that
$r_{\|}^2(0,k_{\bot})=r_{\bot}^2(0,k_{\bot})=1$ as for ideal
metals. In Figs.~\ref{expthBS},\,\ref{expthSL} 
we plot the differences of the Casimir
pressures $\left[P_{1,2}^{\rm theor}-P^{\rm expt}\right]$
versus separation computed
in the approaches of Refs.~\refcite{14},\ \refcite{15} and 
Ref.~\refcite{16}, respectively,
with their respective confidence intervals at 95\% confidence.
As is seen from Fig.~\ref{expthBS}, all points representing
$\left[P_{1}^{\rm theor}-P^{\rm expt}\right]$ obtained
according to Refs.~\refcite{14},\ \refcite{15}
fall outside the confidence interval in a wide separation range from
230\,nm to 500\,nm. (In Fig.~\ref{expthBS} 
the symmetric negative line of the error
bars is not shown because practically all points are positive.)
{}From Fig.~\ref{expthSL} it is seen that almost all points representing
$\left[P_{2}^{\rm theor}-P^{\rm expt}\right]$ obtained according to
Ref.~\refcite{16} fall outside the confidence interval for
separations from 160\,nm to 350\,nm. Thus, the theoretical approaches
of both Refs.~\refcite{14},\ \refcite{15} and \refcite{16} are not
only in contradiction with
thermodynamics but also are excluded experimentally at 95\% confidence.

It is easily seen that for the theory of Refs.~\refcite{14},\ \refcite{15}
the confidence interval can be widened to achieve 99\% confidence
probability. Even in this case almost all differences 
$\left[P_{1}^{\rm theor}-P^{\rm expt}\right]$ fall outside the new
confidence interval in the separation region
$300\,\mbox{nm}\leq z\leq 500\,$nm. We conclude that the theoretical
approach to the thermal Casimir force extensively discussed in
Refs.~\refcite{14},\ \refcite{15},\ \refcite{25a},\ \refcite{25b} is
excluded experimentally with 99\% confidence. This brings a final
resolution to the thermodynamic puzzle arising from the thermal
Casimir force.

\section{Constraints on Hypothetical Long-Range Interactions}

The probable existence of
long-range interactions, in addition to gravitation and
electromagnetism, has long been discussed in elementary particle
physics.\cite{11a} They are predicted by extra-dimensional theories with
low compactification scale,\cite{32} and can also arise from the
exchange of light or massless elementary particles predicted by
many extensions to the Standard Model.\cite{12}
In many cases the effective potential energy between two point masses
$m_1$ and $m_2$ at a distance $r$ can be parametrized by the Newton
gravitational potential with a Yukawa-type correction\cite{33}
\begin{equation}
V(r)=-\frac{Gm_1m_2}{r}\left(1+\alpha_Ge^{-r/\lambda}\right),
\label{eq6}
\end{equation}
\noindent
where $G$ is the gravitational constant, $\alpha_G$ and 
$\lambda$ are the strength and interaction range of the hypothetical
force, respectively.

\begin{figure}[t]
\vspace*{-8.5cm}
\centerline{\psfig{file=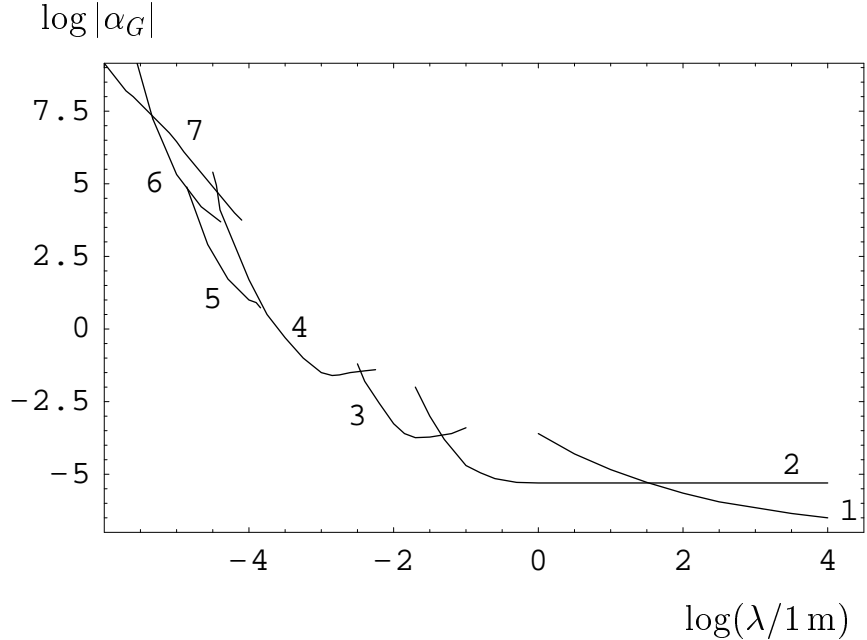,width=20cm}}
\vspace*{-10cm}
\caption{\label{constrG}
Constraints on the strength of the Yukawa-type interaction $\alpha_G$
versus interaction range $\lambda$ obtained from the experiments of
E\"{o}tvos-type (curves 1,\,2),
Cavendish-type (curves 3--6) and
from the measurements of the Casimir force (curve 7).
See text for a more detailed characterization of the curves.
}
\end{figure}

The constraints on corrections to Newtonian gravity follow from
experiments of the E\"{o}tvos- and Cavendish-type. The most stringent
constraints to date are presented in Fig.~\ref{constrG}.
In this figure, the regions of $(\lambda,\alpha_G)$-plane above the curves
are excluded by the results of the indicated experiments, and the
regions below the curves are allowed. Curves 1 and 2 show the
constraints following from the E\"{o}tvos-type experiments of
Refs.~\refcite{34} and \refcite{35}, respectively. Curves 3--6 follow
from the Cavendish-type experiments of Refs.~\refcite{35a}--\refcite{38},
respectively. Constraints on $\alpha_G$ at astronomical scales of
$\lambda$ can be found in Ref.~\refcite{33}.

As is seen from Fig.~\ref{constrG}, 
rather strong constraints on the corrections
to Newtonian gravity are obtained within the interaction range
$\lambda>0.1\,$m. With decreasing $\lambda$ the strength of constraints
falls off rapidly. Curve 7 exhibits the constraints which follow\cite{39}
not from gravitational experiments but from a measurement of the
Casimir force between a plate and a spherical lens by means of a torsion
pendulum.\cite{2} Many other constraints on corrections to Newtonian
gravity in the submicrometer range were obtained from different Casimir
force measurements (see Ref.~\refcite{40} for a review).

As was already noted above, in all previous experiments on the Casimir
effect the root-mean-square deviation $\sigma$ was used as a measure of
agreement between theory and experiment. 
No evidence for hypothetical long-range
interactions has been observed in any experiments.
For this reason the constraints on $\alpha_G,\,\lambda$ were usually
obtained from the inequality
\begin{equation}
|F^{\rm hyp}(z)|\leq\sigma.
\label{eq7}
\end{equation}
\noindent
In doing so it was not possible to quantify the reliability and the
confidence level of the resulting constraints.

The new rigorous approach to the comparison of experiment and theory
in the Casimir force measurements, developed here, gives the
possibility of obtaining constraints on a hypothetical Yukawa-type pressure
from the agreement of our measurements and theory at 95\% confidence.
We have found an agreement between our measurements of the Casimir
pressure and theory in the limits of the error band $\Delta^{\!{\rm tot}}$
calculated at a confidence probability 95\%. Because of this, the
pressure of any hypothetical force must satisfy the inequality
\begin{equation}
| P^{\rm hyp}(z)|\leq\Delta^{\!{\rm tot}}
\left[P^{\rm theor}(z)-P^{\rm expt}(z)\right].
\label{eq8}
\end{equation}
\noindent
It should be stressed that the use of our data to decide among several
competing theories of the thermal Casimir force by no means prevents
us from using the same data to impose stronger constraints on hypothetical
long-range interactions once the choice was made. The reason is that
the Yukawa-type hypothetical pressure depends on separation quite
differently from the alternative thermal corrections discussed above.
Thus, the mimicry of one phenomenon by another one is not possible.
We also note that surface roughness, which was significantly
reduced in this experiment, does not give any noticeable contribution
to a hypothetical interaction of the Yukawa type with an interaction
range of about 100\,nm. Because of this, in calculations of the
hypothetical force one can consider the test bodies to be perfectly
smooth.

\begin{figure}[b]
\vspace*{-8cm}
\centerline{\psfig{file=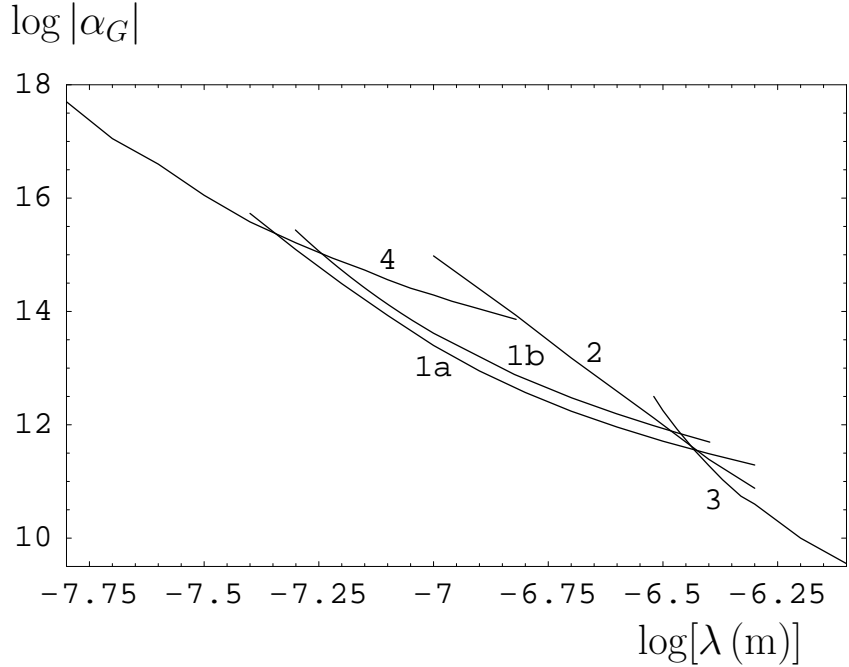,width=20cm}}
\vspace*{-8.2cm}
\caption{\label{constrC}
Constraints on the strength of the Yukawa-type interaction $\alpha_G$
versus interaction range $\lambda$ obtained from the Casimir force
measurements of the present paper (curve 1a) and earlier
measurements of the Casimir force (curves 1b, 2--4).
See text for a more detailed characterization of the curves.
}
\end{figure}
The equivalent hypothetical pressure in the configuration of
two parallel plates 
(one made of Si and coated with Pt and Au layers, the other one made
of Al${}_2$O${}_3$ and coated with Ti and Au layers)
is given by\cite{11a}
\begin{eqnarray}
&&
P^{\rm hyp}(z)=-2\pi G\alpha_G\lambda^2e^{-z/\lambda}
\nonumber \\
&&
\phantom{aaaaaaaa}
\times
\left[\rho_{Au}-\left(\rho_{Au}-\rho_{Ti}\right)e^{-\Delta_{Au}^{\! s}/\lambda}
-\left(\rho_{Ti}-\rho_{AlO}\right)e^{-(\Delta_{Au}^{\! s}+
\Delta_{Ti})/\lambda}
\right]
\label{eq9}\\
&&
\phantom{aaaaaaaa}
\times
\left[\rho_{Au}-\left(\rho_{Au}-\rho_{Pt}\right)e^{-\Delta_{Au}^{\!p}/\lambda}
-\left(\rho_{Pt}-\rho_{Si}\right)e^{-(\Delta_{Au}^{\! p}+
\Delta_{Pt})/\lambda}
\right].
\nonumber
\end{eqnarray}
\noindent
Here the mass densities are given by $\rho_{AlO}=4.1\times 10^3\,$kg/m${}^3$, 
$\rho_{Ti}=4.51\times 10^3\,$kg/m${}^3$,
$\rho_{Au}=19.28\times 10^3\,$kg/m${}^3$,
$\rho_{Si}=2.33\times 10^3\,$kg/m${}^3$,
$\rho_{Pt}=21.47\times 10^3\,$kg/m${}^3$,
and the thicknesses of layers are $\Delta_{Ti}=10\,$nm,
$\Delta_{Au}^{\! s}=200\,$nm,
$\Delta_{Pt}=10\,$nm, 
$\Delta_{Au}^{\! p}=150\,$nm.
Eq.~(\ref{eq9}) was derived under the conditions $z,\,\lambda\ll R,\,L$,
where $L=3.5\mu$m is the thickness of the Si plate.

Constraints obtained from Eq.~(\ref{eq8}) after the substitution of
Eq.~(\ref{eq9}) are plotted by curve 1a in Fig.~\ref{constrC}.
At each $\lambda$
the separation $z$ was found leading to the strongest constraint.
(As a rule, the greater $\lambda$, the greater is $z$ where the 
strongest constraint is obtained.) In the same figure curve 1b shows
constraints obtained from the previous experiment of Ref.~\refcite{11a},
curve 2 --- from an old Casimir force measurement between 
dielectrics,\cite{12}
curve 3 --- from the Casimir force measurement of Ref.~\refcite{2} (this
curve was labeled 7 in Fig.~\ref{constrG}). Curve 4 was obtained\cite{41}
from an experiment measuring the Casimir force by the use of an atomic
force microscope.\cite{6} Note that the constraints of 
curve 1a obtained here
are not only the most stringent ones in the interaction range
$40\,\mbox{nm}\leq\lambda\leq 370\,$nm, but they are also valid at a 95\%
confidence, i.e., with a greater reliability than the other constraints in 
Fig.~\ref{constrC} for which the confidence levels were not determined.
The largest improvement, comparing curves 2 and 4 with 1a, is by
a factor of 20, and is achieved at $\lambda\approx 150\,$nm.

\section{Conclusions and Discussion}

To conclude, we have first experimentally determined the Casimir pressure
between two parallel gold-coated plates with a relative error of
approximately 0.5\% at 95\% confidence within a wide separation range.
The surface
impedance approach to the thermal Casimir force, and two alternative
theoretical approaches advocated in literature, were compared with experiment
in a statistically valid way without recourse to the root-mean-square
deviation. This permitted us to experimentally exclude the alternative
theoretical approaches which predict large thermal
corrections at short separations, and to thus resolve the thermodynamic
puzzle extensively discussed during the past few years. The thermal correction
predicted by the impedance approach to the thermal Casimir force was found
to be consistent with experiment. At $T=300\,$K this correction is small, in
qualitative agreement with the case of ideal metals, and can be readily
measured by means of proposed experiments.\cite{42,43}

Our results lay the groundwork for more precise calculations
of the Casimir forces between closely spaced surfaces, thin films, and
small particles near a cavity wall for applications in nanotechnology,
quantum optics, biology and colloid science.
These results significantly enhance constraints on
predictions of extra-dimensional physics and other extensions to the
Standard Model.
Specifically, constraints on the Yukawa-type hypothetical interaction
were strengthened by a factor of up to 20 within a wide interaction range
at 95\% confidence probability.

\subsection*{Acknowledgments}

The authors thank H.\ B.\ Chan for technical assistance in sample
preparation.
G.\ L.\ Klimchitskaya and V.\ M.\ Mostepanenko are grateful to 
T.\ N.\ Siraya for several helpful
discussions on the metrological procedures of data analysis. They
acknowledge financial support and kind hospitality from  Purdue
University. The work of R.\ S.\ Decca and E.\ Fischbach was 
supported in part by
the Petrolium Research Foundation (through ACS-PRF No.37542--G) and
the U.S.\ Department of Energy (under Contract No.\ DE--AC02--76ER071428),
respectively.

\end{document}